\documentclass[twocolumn,english,aps,showpacs,showkeys,superscriptaddress,citeautoscript]{revtex4}
\usepackage[T1]{fontenc}
\usepackage[latin9]{inputenc}
\usepackage[active]{srcltx}
\usepackage{color}
\usepackage{amsmath}
\usepackage{amssymb}
\usepackage{graphicx}

\makeatletter
\@ifundefined{textcolor}{}
{%
 \definecolor{BLACK}{gray}{0}
 \definecolor{WHITE}{gray}{1}
 \definecolor{RED}{rgb}{1,0,0}
 \definecolor{GREEN}{rgb}{0,1,0}
 \definecolor{BLUE}{rgb}{0,0,1}
 \definecolor{CYAN}{cmyk}{1,0,0,0}
 \definecolor{MAGENTA}{cmyk}{0,1,0,0}
 \definecolor{YELLOW}{cmyk}{0,0,1,0}
}

\include{shadowtext}

\makeatother

\usepackage{babel}
\begin{document}

\title{Quasi-phases and pseudo-transitions in one-dimensional models with
nearest neighbor interactions}

\author{S. M. de Souza }

\affiliation{Departamento de F\'{i}sica, Universidade Federal de Lavras, CP 3037, 37200-000, Lavras-MG, Brazil}

\author{Onofre Rojas}
\email{ors@dfi.ufla.br}

\selectlanguage{english}%

\affiliation{Departamento de F\'{i}sica, Universidade Federal de Lavras, CP 3037, 37200-000, Lavras-MG, Brazil}
\begin{abstract}
There are some particular one-dimensional models, such as the Ising-Heisenberg
spin models with a variety of chain structures, which exhibit unexpected
behaviors quite similar to the first and second order phase transition,
which could be confused naively with an authentic phase transition.
Through the analysis of the first derivative of free energy, such
as entropy, magnetization, and internal energy, a \textquotedbl{}sudden\textquotedbl{}
jump that closely resembles a first-order phase transition at finite
temperature occurs. However, by analyzing the second derivative of
free energy, such as specific heat and magnetic susceptibility at
finite temperature, it behaves quite similarly to a second-order phase
transition exhibiting an astonishingly sharp and fine peak. The correlation
length also confirms the evidence of this pseudo-transition temperature,
where a sharp peak occurs at the pseudo-critical temperature. We also
present the necessary conditions for the emergence of these quasi-phases
and pseudo-transitions.
\end{abstract}


\maketitle
The absence of phase transitions in one-dimensional models with short
range coupling was established since the 1950s, as discussed by van
Hove\cite{Hove}. More recently, Cuesta and Sanchez\cite{cuesta}
investigated relevant properties regarding one-dimensional models,
such as the general non-existence theorem at the finite temperature
 phase transition with short range interaction\cite{dyson-2}. Although
there are some one-dimensional models with long-range interactions
that exhibit phase transition at finite temperature\cite{dyson-1}.
Besides, some peculiar one-dimensional models exhibit at the finite
temperature a first-order phase transition, such as the Kittel model
(zipper model)\cite{kittel}, Chui-Weeks model\cite{chui} and Dauxois-Peyrard
model\cite{dauxois}.

Several real magnetic materials, such as $\mathrm{Cu_{3}(CO_{3})_{2}(OH)_{2}}$
known as natural mineral azurite \cite{kikuchi05}, were investigated
using several approximate methods assuming the Heisenberg model to
describe the natural mineral azurite\cite{Lau}. In addition, Honecker
et al\cite{honecker} investigated the thermodynamic properties of
the Heisenberg model in a diamond chain structure. Furthermore, in
the last decade, the thermodynamics of the Ising-Heisenberg model
in diamond chains has also been widely discussed in references\cite{valverde,Cano,orojas,Lisnii-1}. 

\begin{figure}[h]
\includegraphics[scale=0.7]{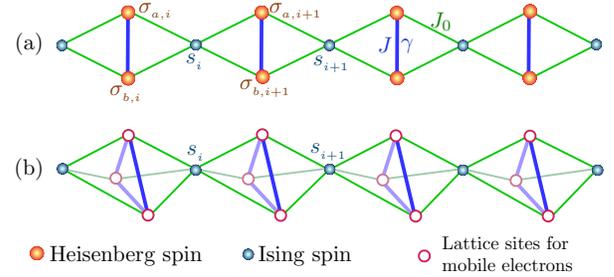}

\caption{\label{fig:chains}Schematic representation: (a) Ising-XYZ diamond
chain model\cite{torrico2}. (b) Double-tetrahedral chain, with localized
Ising spin regularly alternates with two mobile electrons delocalized
over a triangular plaquette\cite{Galisova}.}
\end{figure}

Lately, several one-dimensional models have been investigated in the
framework of decorated structures, particularly Ising and Heisenberg
models with a variety of structures, such as the Ising-Heisenberg
models in diamond chain structure\cite{torrico,torrico2} as shown
in fig.\ref{fig:chains}a, one-dimensional double-tetrahedral chain
(see fig.\ref{fig:chains}b), in which the localized Ising spin regularly
alternates with two mobile electrons delocalized over a triangular
plaquette\cite{Galisova}, alternating Ising-Heisenberg ladder model\cite{on-strk},
Ising-Heisenberg triangular tube model\cite{strk-cav}. The analysis
of the first derivative of the thermodynamic potential, such as entropy,
internal energy, magnetization shows a significant jump as a function
of temperature, maintaining a close similarity with the first order
phase transition. Similarly, a second order derivative of potential
thermodynamics, such as specific heat and magnetic susceptibility,
resembles a typical second order phase transition at finite temperature.

\paragraph{Quasi-phases and pseudo-transitions:}

Most one-dimensional models with short-range interaction have been
extensively investigated in the last decade\cite{valverde,Cano,orojas,Lisnii-1,torrico,torrico2,Galisova},
whose transfer matrix have the following structure $\mathbf{T}=\left[\begin{array}{cc}
w_{0} & w_{1}\\
w_{1} & w_{2}
\end{array}\right]$. Obviously, the corresponding eigenvalues are given by $\lambda_{\pm}=\tfrac{1}{2}\Bigl(w_{0}+w_{2}\pm\sqrt{(w_{0}-w_{2})^{2}+4w_{1}^{2}}\Bigr)$,
where $w_{n}=\sum_{k=0}{\rm e}^{-\beta\varepsilon_{n,k}}$ are the
Boltzmann factors, with $\varepsilon_{n,k}$ being the energy levels
$k=\{0,1,\ldots\}$ for each sector $n=0$, 1 and 2. For simplicity,
here we are considering only the non-degenerate case since its extension
to the degenerate case is trivial. Where $\beta=1/k_{B}T$ with $k_{B}$
Boltzmann constant and $T$ the absolute temperature.

Therefore, for a periodic chain with $N$ unit cell, the partition
function can be expressed by $\mathcal{Z}_{N}=\lambda_{+}^{N}+\lambda_{-}^{N}$,
while the free energy per unit cell in thermodynamic limit becomes
\begin{equation}
f=-\frac{1}{\beta}\ln\left[\tfrac{1}{2}\Bigl(w_{0}+w_{2}+\sqrt{(w_{0}-w_{2})^{2}+4w_{1}^{2}}\Bigr)\right].\label{eq:f-ex}
\end{equation}

Now let us ask the following question, what happens to the free energy
$f$ when $w_{1}=0$? This means that free energy can be described
as $f=-\frac{1}{\beta}\ln(w_{0})$ for $w_{0}>w_{2}$, whereas $f=-\frac{1}{\beta}\ln(w_{2})$
for $w_{0}<w_{2}$, or simply expressed by $f=\min\bigl\{-\frac{1}{\beta}\ln(w_{0}),-\frac{1}{\beta}\ln(w_{2})\bigr\}$,
which is piecewise function. On the other hand, the
transfer matrix $\mathbf{T}$ becomes diagonal matrix with elements
$w_{0}$ and $w_{2}$. The limit $w_{0}=w_{2}$ leads to a transcendental
equation involving the temperature, and this one can be solved numerically,
from which we can find a genuine critical temperature. 

In particular the Hamiltonian for the one-dimensional Ising model
with spin-1/2 is $\mathcal{H}=\sum_{i}(-Js_{i}s_{i+1}-hs_{i})$. The
energies per unit cell are $\varepsilon_{0,0}=-\frac{J}{4}-\frac{h}{2}$
for ($s_{i},s_{i+1}$) $\rightarrow$ ($\uparrow,\uparrow$), $\varepsilon_{2,0}=-\frac{J}{4}+\frac{h}{2}$
for ($s_{i},s_{i+1}$) $\rightarrow$ ($\downarrow,\downarrow$),
and $\varepsilon_{1,0}=\frac{J}{4}$ for ($s_{i},s_{i+1}$) $\rightarrow$
($\uparrow,\downarrow$) or ($\downarrow,\uparrow$), thus the Boltzmann
factors simply reduce to: $w_{0}={\rm e}^{-\beta\varepsilon_{0,0}}$,
$w_{1}={\rm e}^{-\beta\varepsilon_{1,0}}$ and $w_{2}={\rm e}^{-\beta\varepsilon_{2,0}}$.
Despite the above condition $w_{1}\rightarrow0$ occurs when $J\rightarrow\infty$,
we will never have the competing condition between $w_{0}$ and $w_{2}$
for $h>0$. Besides, the condition $w_{0}=w_{2}$ for the Ising one-dimensional
Ising model just implies $\varepsilon_{0,0}=\varepsilon_{2,0}$ and
obviously there is no non-zero \textquotedbl{}critical temperature\textquotedbl{}.

But is it possible that $w_{1}=0$? Note that $w_{1}=\sum_{r=0}{\rm e}^{-\beta\varepsilon_{1,r}}$.
So we conclude that we will never have $w_{1}=0$, for $\beta<\infty$.
Thus from now on, we will rigorously analyze the case $w_{1}>0$.

Several \textquotedbl{}decorated\textquotedbl{} one-dimensional models\cite{valverde,Cano,orojas,Lisnii-1}
(and references there in) satisfy the following condition $w_{0}\sim w_{2}\sim w_{1}$
and $|w_{0}-w_{2}|\sim2w_{1}$, with well-known thermodynamic properties.
However, there are some particular models\cite{torrico2,Galisova,on-strk,strk-cav}
that satisfy the following condition $w_{0}\sim w_{2}\ggg w_{1}$
and $|w_{0}-w_{2}|\gg2w_{1}$. The free energy using the Taylor series
expansion around $w_{1}\rightarrow0$, results in
\begin{equation}
f\approx\begin{cases}
-\tfrac{1}{\beta}\ln(w_{0})-\tfrac{1}{\beta}\tfrac{w_{1}^{2}}{w_{0}^{2}}, & w_{0}\geqslant w_{2}\\
-\tfrac{1}{\beta}\ln(w_{2})-\tfrac{1}{\beta}\tfrac{w_{1}^{2}}{w_{2}^{2}}, & w_{0}<w_{2}
\end{cases}.\label{f-apx}
\end{equation}

\begin{figure}[h]
\includegraphics[scale=0.44]{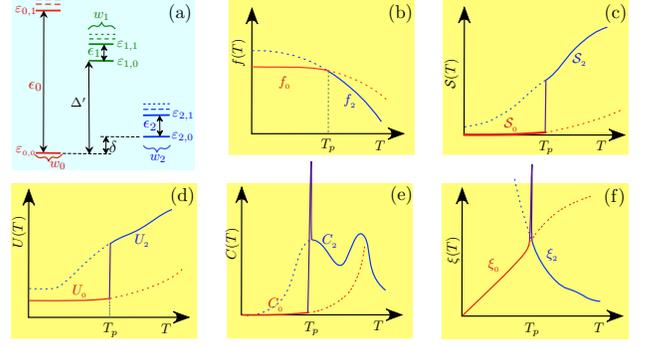}\caption{\label{fig:Free-energy}(a) Low-lying energy levels. (b) Free energy,
(c) Entropy, (d) Internal energy, (e) Specific heat, and (f) correlation
length as a function of temperature. With $T_{p}$ being the pseudo
critical temperature. }
\end{figure}

To analyze the most relevant behavior of \eqref{eq:f-ex} without
losing its generality, we consider only a couple of lowest energies
for each Boltzmann factors (as described in the figure \ref{fig:Free-energy}a),
using this feature each sector can be expressed as: For ($n=0$),
the Boltzmann factor reduce to $w_{0}={\rm e}^{-\beta\varepsilon_{0,0}}+{\rm e}^{-\beta\varepsilon_{0,1}}$
with $\varepsilon_{0,0}<\varepsilon_{0,1}$. For ($n=1$) the $w_{1}={\rm e}^{-\beta\varepsilon_{1,0}}+{\rm e}^{-\beta\varepsilon_{1,1}}$
depends of the low-lying excited energies $\varepsilon_{1,0}$ and
$\varepsilon_{1,1}$. Whereas, for ($n=2$) the $w_{2}={\rm e}^{-\beta\varepsilon_{2,0}}+{\rm e}^{-\beta\varepsilon_{2,1}}$
are given in terms of $\varepsilon_{2,0}$ and $\varepsilon_{2,1}$. 

The energy gaps outlined in fig.\ref{fig:Free-energy}a are $\epsilon_{0}=\varepsilon_{0,1}-\varepsilon_{0,0}$,
$\epsilon_{1}=\varepsilon_{1,1}-\varepsilon_{1,0}$, $\epsilon_{2}=\varepsilon_{2,1}-\varepsilon_{2,0}$,
$\Delta'=\varepsilon_{1,0}-\varepsilon_{0,0}$ and $\delta=\varepsilon_{2,0}-\varepsilon_{0,0}$.
The above condition of $w_{n}$'s in terms of energy gaps become roughly
$\delta\sim\epsilon_{2}\ll2\Delta'\lesssim\epsilon_{0}$, and define
conveniently $\Delta={\rm min}(\Delta',\epsilon_{0}/2)$.

In fig.\ref{fig:Free-energy}b is illustrated the free energy $f_{0}\approx\varepsilon_{0,0}-T\,{\rm e}^{-\frac{2\Delta}{T}}$
when $w_{0}>w_{2}$, which is weakly dependent of other energy levels,
because the main contribution of the free energy is $\varepsilon_{0,0}$.
While for $w_{0}<w_{2}$, the free energy $f_{2}\approx\varepsilon_{2,0}-T\,\ln\left(1+{\rm e}^{-\epsilon_{2}/T}\right)$
depends more significantly of $\epsilon_{2}$. Although
free energy resembles a piecewise function, exact free energy\eqref{eq:f-ex}
is an analytic function or an infinitely differentiable function.

When $w_{1}\ll1$, the transfer matrix $\mathbf{T}$ is a quasi diagonal
matrix, with competing $w_{0}$ and $w_{2}$, so the condition $w_{0}=w_{2}$
lead us to a transcendental equation in temperature or any other Hamiltonian
parameter, so this equation can be solved numerically to find a \textquotedbl{}pseudo-critical\textquotedbl{}
temperature $T_{p}$. 

In fig.\ref{fig:Free-energy}c is depicted schematically the entropy
$\mathcal{S}(T)$ as a function of the temperature, for $w_{0}>w_{2}$
or $T<T_{p}$ we observe the entropy would be almost a constant curve
$\mathcal{S}_{0}\approx\left(1+\frac{2\Delta}{T}\right){\rm e}^{-2\Delta/T}$,
whereas the entropy \textquotedbl{}suddenly\textquotedbl{} increases
at $T_{p}$ and for $w_{0}\leqslant w_{2}$ or $T_{p}\leqslant T$
it increases significantly as soon the temperature increases with
$\mathcal{S}_{2}\approx\ln(1+{\rm e}^{-\epsilon_{2}/T})+\frac{\epsilon_{2}}{T}\left(1+{\rm e}^{\epsilon_{2}/T}\right)^{-1}$.
Indeed, some  evidence of this behavior has already been observed
in previous work\cite{Galisova,on-strk,strk-cav,timonin}, and spin-ice
model with short-range interaction in the Bethe-Peierls approach\cite{timonin}.
The entropy behavior reveals the similarity with the first-order phase
transition\cite{Sauera}, so it is interesting to define the \textquotedbl{}latent
heat\textquotedbl{} associated with the system in $T_{p}$ as $\mathcal{L}\propto Q\equiv T_{p}\left(\mathcal{S}_{2}(T_{p})-\mathcal{S}_{0}(T_{p})\right)=T_{p}\varDelta\mathcal{S}_{2,0}$.

The internal energy $U(T)$ is also depicted schematically in fig.\ref{fig:Free-energy}d,
where we can observe a nearly constant energy $U_{0}\approx\varepsilon_{0,0}+2\Delta{\rm e}^{-2\Delta/T}$
for $T<T_{p}$, whereas for $T_{p}<T$ the thermal excitation significantly
influences the internal energy $U_{2}\approx\varepsilon_{2,0}+\epsilon_{2}\left(1+{\rm e}^{\epsilon_{2}/T}\right)^{-1}$
like in most one-dimensional models. Again the jump in internal energy
shows the similarity to the first order phase transition\cite{Sauera},
so alternatively we can express the \textquotedbl{}latent heat\textquotedbl{}
as $\mathcal{L}=Q=U_{2}(T_{p})-U_{0}(T_{p})=\delta+\epsilon_{2}(1+{\rm e}^{\epsilon_{2}/T_{p}})^{-1}-2\Delta{\rm e}^{-2\Delta/T_{p}}$,
this quantity reinforces the first-order pseudo-transition. 

However, since the free energy is an analytic function, the second
derivative of the free energy near the pseudo-transition temperature
exhibits an impressively fine peak quite similar to a cusp-like singularity
of second-order phase transition, as occurs in magnetic susceptibility
and specific heat (see fig.\ref{fig:Free-energy}e). Near the pseudo-critical
temperature, the specific heat becomes $C_{0}\approx\left(\tfrac{2\Delta}{T}\right)^{2}{\rm e}^{-2\Delta/T}$
for $T<T_{p}$ and $C_{2}\approx\left(\tfrac{\epsilon_{2}}{T}\right)^{2}{\rm e}^{-\epsilon_{2}/T}\left(1+{\rm e}^{-\epsilon_{2}/T}\right)^{-2}$
for $T_{p}<T$. Some evidence of this effect has already been observed
in one-dimensional models such as those discussed in the literature\cite{Galisova,on-strk,strk-cav}. 

It is also worth investigating the correlation length $\xi$ illustrated
in fig.\ref{fig:Free-energy}f (i.e. nodal Ising spin correlation
length), which is given by $\xi^{-1}=\ln\left(\tfrac{\lambda_{+}}{\lambda_{-}}\right)$.
Using the Taylor series expansion analogous to eq.\eqref{f-apx} when
$w_{1}\rightarrow0$, the correlation length becomes, 
\begin{alignat}{1}
\xi^{-1}\approx & \begin{cases}
\ln\left(\tfrac{w_{0}}{w_{2}}+\tfrac{w_{1}^{2}}{w_{2}^{2}}\right), & w_{0}>w_{2}\\
\ln\left(\tfrac{w_{2}}{w_{0}}+\tfrac{w_{1}^{2}}{w_{0}^{2}}\right), & w_{0}<w_{2}
\end{cases}.
\end{alignat}
 The leading terms of the correlation length can be expressed by $\xi_{0}\approx\left\{ \frac{\delta}{T}-\ln\left(1+{\rm e}^{-\epsilon_{2}/T}\right)\right\} ^{-1}$
for $T<T_{p}$, and $\xi_{2}\approx\left\{ -\frac{\delta}{T}+\ln\left(1+{\rm e}^{-\epsilon_{2}/T}\right)\right\} ^{-1}$
for $T_{p}<T$. Of course, this expression is valid around the pseudo-critical
temperature, and this expression give us simply a maximum in $T=T_{p}$,
and not a cusp-like singularity.

The present analysis could be applied to any physical system that
has the form of free energy \eqref{eq:f-ex}. Here, we only discuss
how the unusual properties arise near the pseudo-critical temperature
$T_{p}$.

Now let us consider as an illustrative example the one-dimensional
Ising-XYZ model in the diamond chain structure which was solved exactly
in the references\cite{torrico,torrico2}.

\paragraph{Ising-XYZ diamond chain:}

In fig.\ref{fig:chains}a, the Ising-XYZ diamond chain structure is
schematically illustrated. Where $s_{i}$ represents the Ising spin-1/2,
and $\sigma_{a(b),i}^{\alpha}$ denotes the Heisenberg spin-1/2, assuming
$\alpha=\{x,y,z\}$, whose Hamiltonian and its exact solution was
given in reference \cite{torrico2}.

The couples of low-lying energy levels for each sector $w_{n}$ are:

(i) For sector $n=2$ ($\uparrow\uparrow$ ) the first low-lying energy
is 
\begin{equation}
\varepsilon_{2,0}=E_{_{MF_{2}}}=-\tfrac{J_{z}}{4}-\tfrac{h}{2}-\sqrt{(h+J_{0})^{2}+\tfrac{1}{4}J^{2}\gamma^{2}},
\end{equation}
(ferromagnetic Ising spin and modulated ferromagnetic Heisenberg spin
($MF_{2}$) phase), and the other energy level is 
\begin{equation}
\varepsilon_{2,1}=E_{_{FI}}=-\tfrac{J+h}{2}+\tfrac{J_{z}}{4},
\end{equation}
(ferrimagnetic (FI) phase).

The corresponding ground states are expressed by 
\begin{alignat}{1}
|MF_{n}\rangle= & \overset{N}{\underset{i=1}{\prod}}c_{n}\left(\alpha_{n}|\begin{smallmatrix}+\\
+
\end{smallmatrix}\rangle_{i}+|\begin{smallmatrix}-\\
-
\end{smallmatrix}\rangle_{i}\right)\otimes|\tau\rangle_{i},\label{eq:FMF1}\\
|FI\rangle= & \overset{N}{\underset{i=1}{\prod}}\tfrac{1}{\sqrt{2}}\left(|\begin{smallmatrix}-\\
+
\end{smallmatrix}\rangle_{i}+|\begin{smallmatrix}+\\
-
\end{smallmatrix}\rangle_{i}\right)\otimes|+\rangle_{i},
\end{alignat}
where $\alpha_{n}=\frac{-J\gamma}{2h+2J_{0}\mu-2A_{n}}$, $c_{n}=\frac{1}{\sqrt{1+\alpha_{n}^{2}}}$
and $A_{n}=\sqrt{\left(h+J_{0}(n-1)\right)^{2}+\tfrac{1}{4}J^{2}\gamma^{2}}$.
The state $MF_{2}$ is obtained fixing $n=2$ and $\tau=+$.

(ii) Whereas in sector $n=0$ ($\downarrow\downarrow$), we have the
ground state energy, whose energy becomes 
\begin{equation}
\varepsilon_{0,0}=E_{_{MF_{0}}}=-\tfrac{J_{z}}{4}+\tfrac{h}{2}-\sqrt{(h-J_{0})^{2}+\tfrac{1}{4}J^{2}\gamma^{2}}.
\end{equation}

The corresponding modulated ferromagnetic ($MF_{0}$) state is given
by \eqref{eq:FMF1} when $n=0$ and $\tau=-$. Whereas, the first
excited energy in this sector is $\varepsilon_{0,1}=\frac{J_{z}}{4}-\frac{J-h}{2}$.

(iii) Analogously, for sector $n=1$ ($\downarrow\uparrow$ or $\uparrow\downarrow$)
a couple of low-lying excited energy levels are $\varepsilon_{1,0}=\frac{J_{z}}{4}-\frac{J}{2}$
and $\varepsilon_{1,1}=-\frac{J_{z}}{4}-\frac{1}{2}\sqrt{4h^{2}+J^{2}\gamma^{2}}$.

\paragraph{\textquotedbl{}Quasi-phase\textquotedbl{} diagram for Ising-XYZ diamond
chain:}

\begin{figure}[h]
\includegraphics[scale=0.48]{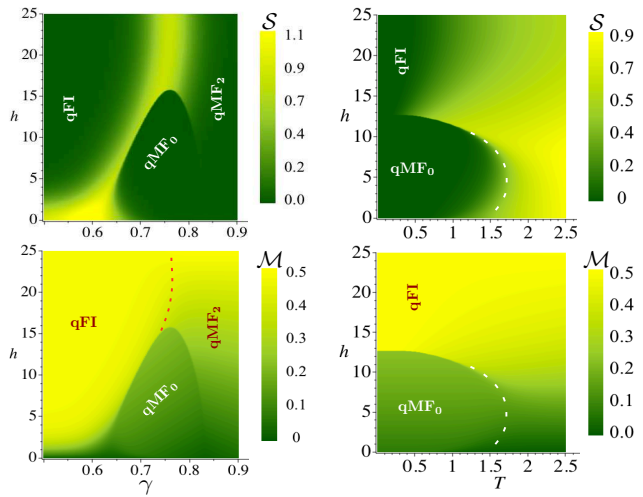}\caption{Density plot: \label{fig:QPh-dmd}Quasi-phase diagram for Ising-XYZ
diamond chain at finite temperature, for fixed $J=100$, $J_{0}=-24$
and $J_{z}=24$. (Left) for fixed $T=1$, (right) for fixed $\gamma=0.7$.}
\end{figure}

All plots below were performed using the exact result \cite{torrico2},
while the low temperature limit discussed above fits nicely with the
exact result.

In fig.\ref{fig:QPh-dmd}(left-top) the density plot for the entropy
in the $\gamma-h$ plane is shown for the parameters given in the
legend of fig.\ref{fig:QPh-dmd}, this plot resembles the vestiges
of the phase diagram at zero temperature\cite{torrico2}. Then due
to the thermal excitation, these phases will be called as \textquotedbl{}quasi-phase\textquotedbl{},
the boundary between $qFI$-$qMF_{2}$ ($FI$-$MF_{2}$ at $T=0$)
there is a standard phase transition signaling that becomes smooth
due to thermal excitation. However, the boundary between $qFI$-$qMF_{0}$
($FI$-$MF_{0}$ ) and $qMF_{0}$ -$qMF_{2}$ ($MF_{0}$ -$MF_{2}$
) exhibits an uncommonly well-defined boundary around the $qMF_{0}$
region, and this region seems insensitive to thermal excitation, this
is because in this region there is a large energy gap $2\Delta$.
This phenomenon is very unusual for one-dimensional models, because
any traces of zero temperature must be fade away as the temperature
increases. A similar plot is depicted for the magnetization in fig.\ref{fig:QPh-dmd}(left-bottom),
the density plot illustrates the magnetization for the same set of
entropy parameters. We can observe the boundary between $qFI$-$qMF_{2}$
is almost imperceptible, here is marked by dashed line just to follow
the phase transition pattern. However, the boundary between $qFI$-$qMF_{0}$
and $qMF_{0}$ -$qMF_{2}$ is completely different with clearly sharp
boundary, rounding the $qMF_{0}$ region. 

Now let us show in fig.\ref{fig:QPh-dmd}(right) the quasi-phase diagram
in the $h$-$T$ plane, the density plot for entropy (top) and magnetization
(bottom). For most one-dimensional models, this type of diagram will
only show traces of phase transition at zero temperature which readily
disappears when the temperature increases. However, here the entropy/magnetization
density plot illustrates that the sharp boundary clearly survives
at finite temperature, and seems almost independent of thermal excitation,
although at a higher temperature that sharp boundary vanishes. 

\begin{figure}[h]
\includegraphics[scale=0.15]{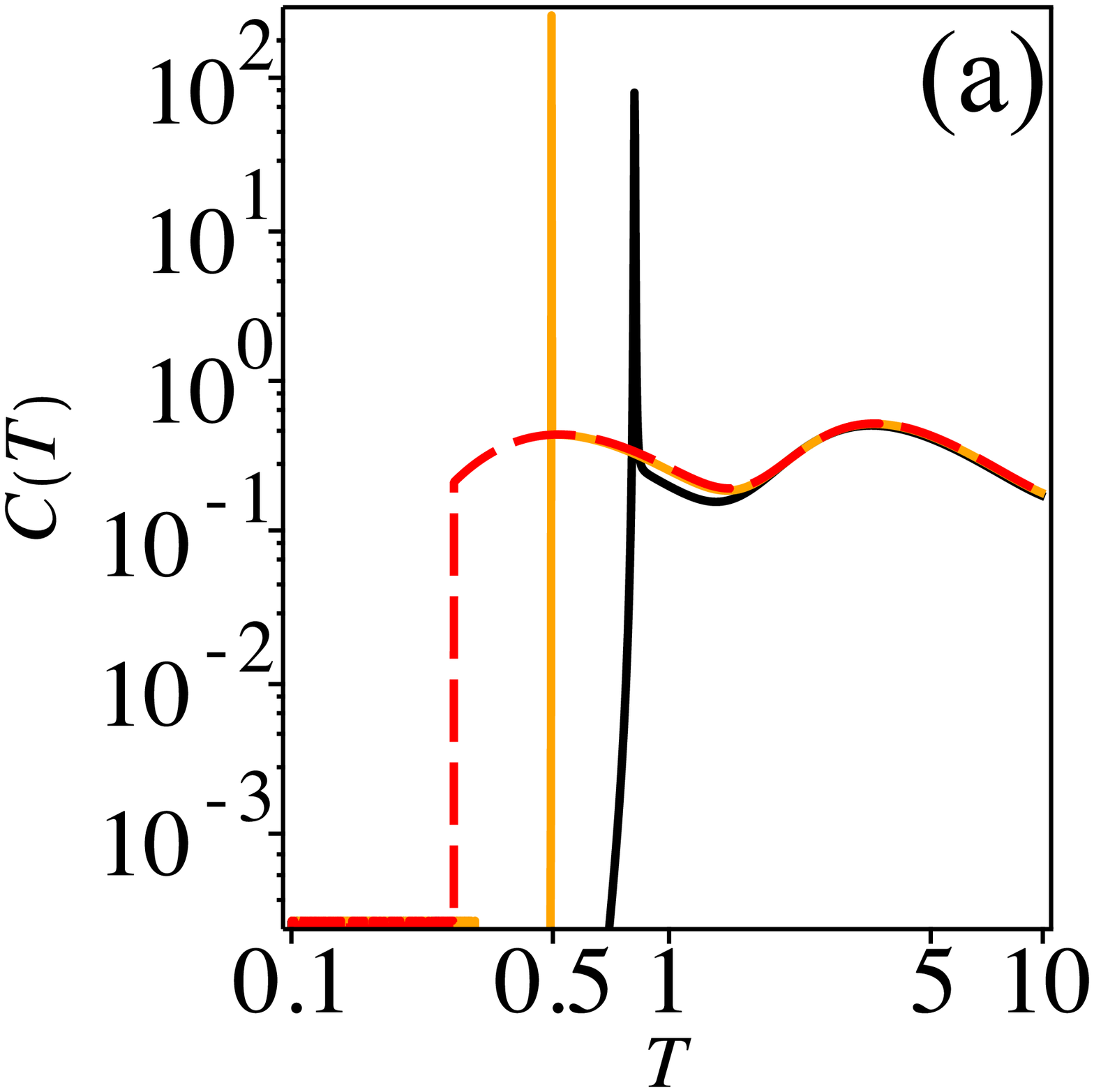}\includegraphics[scale=0.15]{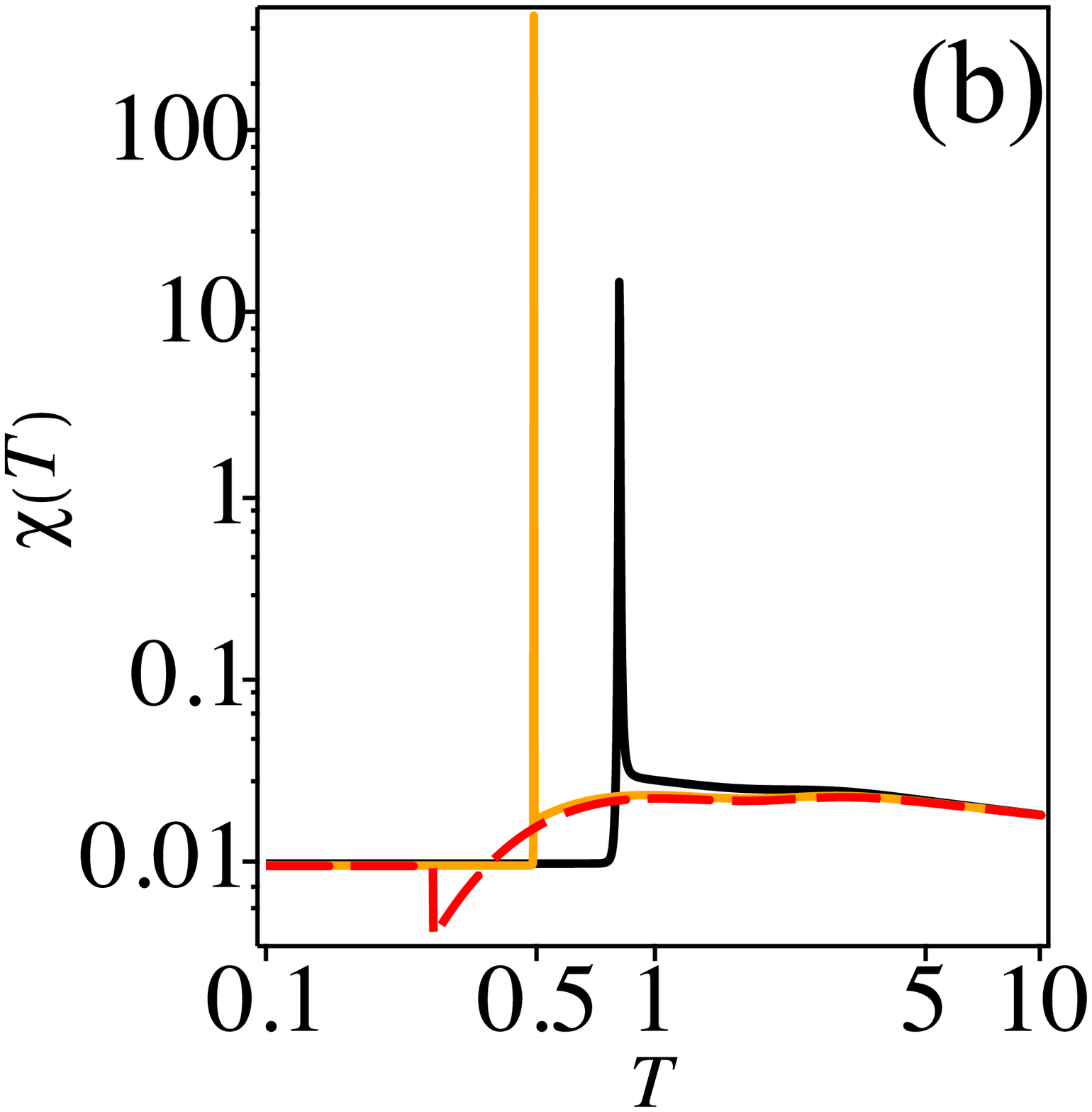}\includegraphics[scale=0.15]{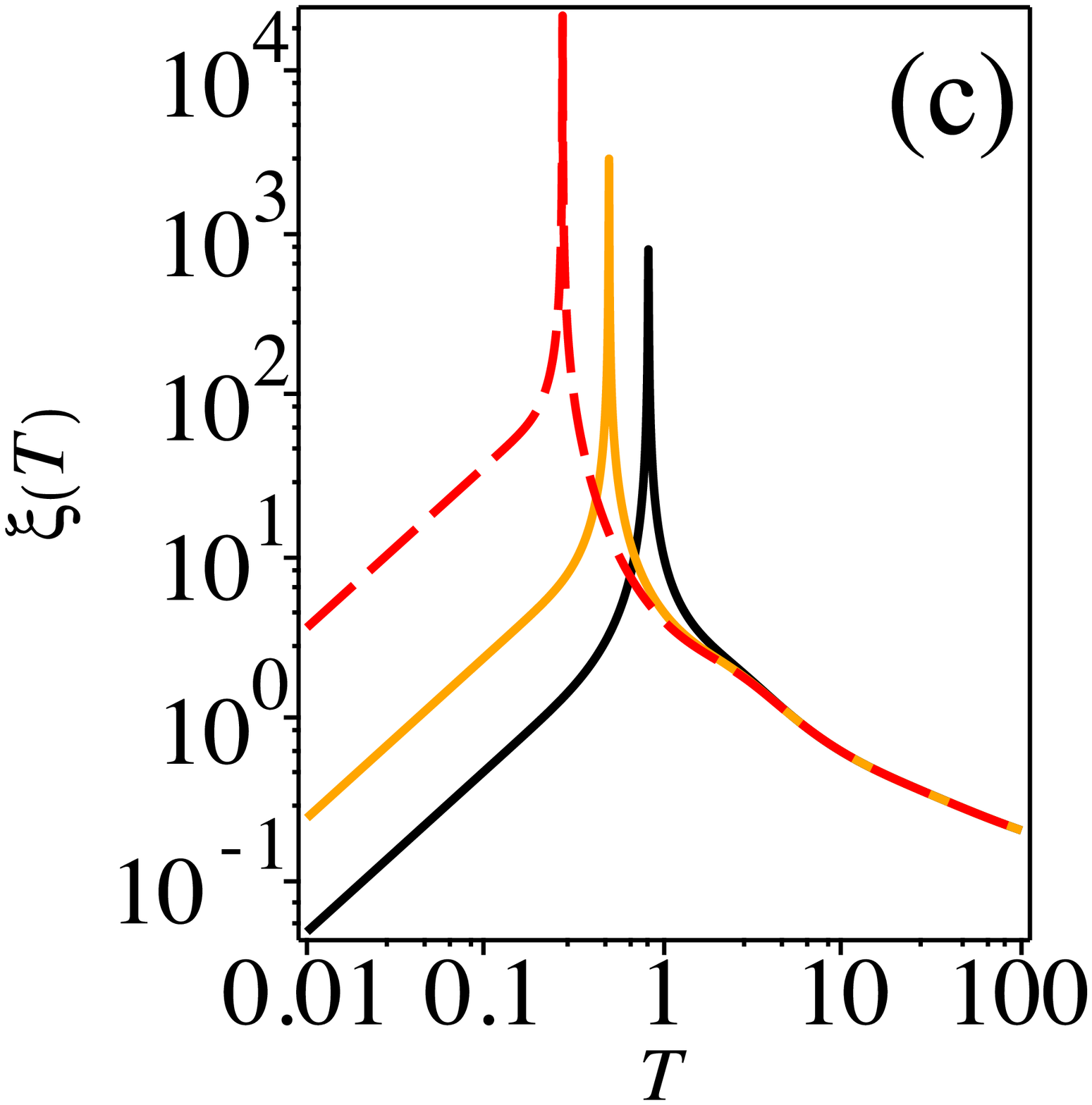}\caption{\label{fig:pseudo-Tp}Ising-XYZ diamond chain\cite{torrico,torrico2}:
(a) Specific heat as a function of temperature temperature. (b) Magnetic
susceptibility. In (c) correlation length $\xi(T)$ as a function
of temperature. Assuming fixed parameters $J=100$, $J_{0}=-24$ and
$J_{z}=24$ and $\gamma=0.7$ for $h=12$ (black line), $h=12.6$
(orange line) and $h=12.74$ (red dashed line). }
\end{figure}

In fig.\ref{fig:pseudo-Tp} is illustrated the pseudo-transition temperature
for the Ising-XYZ diamond chain\cite{torrico2}, where an extremely
strong fine peak occurs for specific heat (a), magnetic susceptibility
(b) and correlation length between nodal Ising spins(c). All panels
are conveniently drawn on logarithmic scales. It is worth mentioning
that this peak never goes to infinity, the smaller the $T_{p}$ the
thinner and stronger the peak becomes. Note that in panel (a) and
(b) the red dashed line apparently shows the absence of peaks, but
there is an astonishingly thin and vigorous peak. Indeed, only for
$T_{p}=0$ the peak leads to infinity indicating a genuine phase transition,
which is in agreement with the phase transition theorem\cite{cuesta}.

\paragraph{Conclusions.}

Although, there are no real phase transitions in the one-dimensional
model. For some special cases of Ising-Heisenberg one-dimensional
spin models, the analysis of internal energy, entropy and magnetization
show a pseudo-transition quite similar to a first-order phase transition,
where we associate a \textquotedbl{}latent heat\textquotedbl{} to
reinforce this property\cite{Sauera}. While for specific heat and
magnetic susceptibility it exhibits an extremely sharp peak indicating
a \textquotedbl{}pseudo-transition\textquotedbl{} between the quasi-phases,
this pseudo-transition closely resembles a typical second-order phase
transition. Some evidence of quasi-phase and pseudo-transitions have
already been manifested in recent previous works, such as Ising-XYZ
diamond chain\cite{torrico2}, tetrahedral chain\cite{Galisova},
Ising-Heisenberg ladder model\cite{on-strk}, Ising-Heisenberg triangular
tube models\cite{strk-cav}. However, here we present a general condition
for appearing this quasi-phases and pseudo-transitions. It is worth
mentioning that this unexpected property is intrinsically related
to the \textquotedbl{}decorated\textquotedbl{} lattice models. Evidently,
this opens several possibilities of finding  other \textquotedbl{}decorated\textquotedbl{}
model with this property. Another relevant issue to note is that this
result opens the possibility of researching and synthesizing real
materials with this stunning property.

This work was partially supported by Brazilian agencies CNPq and FAPEMIG.

\end{document}